# Initial and Eventual Software Quality Relating to Continuous Integration in GitHub

Yue Yu‡†, Bogdan Vasilescu†, Huaimin Wang‡, Vladimir Filkov†, Premkumar Devanbu†

‡College of Computer
National University of Defense Technology
Changsha, 410073, China
{yuyue, hmwang}@nudt.edu.cn

†Department of Computer Science
University of California, Davis
Davis, CA 95616, USA
{vasilescu, ptdevanbu, vfilkov}@ucdavis.edu

## ABSTRACT

The constant demand for new features and bug fixes are forcing software projects to shorten cycles and deliver updates ever faster, while sustaining software quality. The availability of inexpensive, virtualized, cloud-computing has helped shorten schedules, by enabling *continuous integration* ($\mathcal{CI}$) on demand. Platforms like GITHUB support $\mathcal{CI}$ in-the-cloud. In projects using $\mathcal{CI}$, a user submitting a pull request triggers a $\mathcal{CI}$ step. Besides speeding up build and test, this fortuitously creates voluminous archives of build and test successes and failures. $\mathcal{CI}$ is a relatively new phenomenon, and these archives allow a detailed study of $\mathcal{CI}$. How many problems are exposed? Where do they occur? What factors affect $\mathcal{CI}$ failures? Does the "initial quality" as ascertained by $\mathcal{CI}$ predict how many bugs will later appear ("eventual quality") in the code? In this paper, we undertake a large-scale, fine resolution study of these records, to better understand $\mathcal{CI}$ processes, the nature, and predictors of $\mathcal{CI}$ failures, and the relationship of $\mathcal{CI}$ failures to the eventual quality of the code. We find that: a) $\mathcal{CI}$ failures appear to be concentrated in a few files, just like normal bugs; b) $\mathcal{CI}$ failures are not very highly correlated with eventual failures; c) The use of $\mathcal{CI}$ in a pull request doesn't necessarily mean the code in that request is of good quality.

## 1. INTRODUCTION

Modern software projects are under increasing pressures to deliver new features and bug fixes rapidly. These pressures arise from greater competition, the rise of 0-day exploits, and increasing customer expectations for new features. In addition, the increasing popularity of cloud-based services allows changes to be deployed very quickly indeed since there is no need for customers to receive shrink-wrapped updates, or even download anything. Instead, new code need only be installed on the cloud for customers to experience it. Expected cycle-times for novelty in products have gone down from years to months to weeks, to in some cases days or even hours.

A key enabler of rapid-fire delivery is the ability to build and test changes quickly, and thoroughly. Build and test procedures can ravenously consume resources; in the past, these steps were undertaken sparingly and cautiously. But now, thanks to cloud-enabled distributed development, projects can acquire and dispose computing resources with high dynamics; in turn, this enables extremely rapid build and test cycles, with the possibility of commandeering thousands of CPUs as and when needed. Cloud-based *Continuous Integration* ($\mathcal{CI}$) [11] is a rapidly growing approach in many popular open-source and commercial projects. Tools such as TRAVIS-CI[1] are used to frequently build code and run tests (including unit, integration, and system tests) on new changes. This approach, in theory, sustains a level of quality in the code as new features and bug fixes are added.

Open-source projects following GITHUB's pull-based development model have another incentive to adopt $\mathcal{CI}$: the ability to manage large volumes of contributions in the form of *Pull Request* ($\mathcal{PR}$), coming from many people. $\mathcal{CI}$ provides a way to automatically and rapidly vet contributions to ensure that they build, and pass all the available unit, integration, and regression tests. The $\mathcal{CI}$ process provides a "quality gate" for each $\mathcal{PR}$; any build or test failures found during $\mathcal{CI}$ are recorded, and typically *must* be fixed before the $\mathcal{PR}$ can be merged. Once any found defects are fixed, and if subsequent code review is positive and the $\mathcal{PR}$ contribution is deemed of sufficient value, it may be accepted. We describe the process in more detail in Section 2.

But is $\mathcal{CI}$ working? Is it really helping to find and fix potential quality problems (if any) with submitted pull requests? That is, do $\mathcal{PR}$s with $\mathcal{CI}$-identified failures, and subsequent fixes during $\mathcal{PR}$ submission, still continue to have quality issues after the $\mathcal{PR}$ is merged? If $\mathcal{CI}$ is working really well, it will expose most of the defects in $\mathcal{PR}$s prior to merging, so that *after* the merge the new code will work reliably, even if it had initial problems.

The crucial aspect of relevance here is that *all build and test failures exposed during $\mathcal{CI}$ are logged*, providing a permanent record of initial quality. Automated build and testing of code is not new; it has been going on since the early days of software engineering. What is new here is the abundance of data concerning $\mathcal{CI}$, from platforms such as GITHUB. This data provides a new opportunity to evaluate how well $\mathcal{CI}$ is working. In this paper, *we focus on all failures uncovered by the $\mathcal{CI}$ process, in both build & test*.

Designers of automated $\mathcal{CI}$ scripts are certainly not solely concerned with velocity, but also with thoroughness. A good

---
[1] https://travis-ci.org/

$\mathcal{CI}$ testing script is an attempt to expose as many defects as possible, before merging, so that merged code will be relatively defect-free. The $\mathcal{CI}$ log of a $\mathcal{PR}$, then, represents its initial, *pre-merge* defect record, which we call *initial quality record* ($\mathcal{IQR}$). After initial defects are fixed, and if the $\mathcal{PR}$ is merged, it will be used by customers, who may also find failures, which are repaired; these repairs constitute the *post-merge* defect record, which we call *eventual quality record* ($\mathcal{EQR}$), of that specific code. Both these phenomena (initial failures, and changes made to fix bugs discovered post-integration) can be observed: the former in $\mathcal{CI}$ logs, and the latter using well-known procedures to identify bug-fix commits, and the fix-inducing code in version control systems.

Given the large number and variety of projects in GITHUB, we have a new opportunity to study both the initial quality, $\mathcal{IQR}$, and it's relationship to the eventual quality, $\mathcal{EQR}$. In this paper, we undertake extensive mining of $\mathcal{PR}$ histories in many large projects. We mine the $\mathcal{IQR}$ of the $\mathcal{PR}$s, as observed from TRAVIS-CI logs. We also mine the $\mathcal{EQR}$ of these $\mathcal{PR}$s, by observing how post-merge defect repair can be mapped to these $\mathcal{PR}$s. These are some of the most noteworthy contributions of this paper:

- It is well known that a few files account for most bugs [29] in software systems. We find that the pre-merge $\mathcal{CI}$ failures are *just as concentrated* into a few files.
- We find that $\mathcal{PR}$s affecting files with a history of $\mathcal{CI}$ failures are associated with more $\mathcal{CI}$ failures later.
- We find that $\mathcal{PR}$s touching more frequently changed files, surprisingly, have fewer $\mathcal{CI}$ failures.
- We find that $\mathcal{PR}$s touching *files* with a worse $\mathcal{IQR}$ (more $\mathcal{CI}$ failures) don't necessarily have worse $\mathcal{EQR}$.
- However, we did find that the code originating in $\mathcal{PR}$s with failures in $\mathcal{IQR}$ has 31.5% greater odds of being associated with an eventual quality problem.

Our findings suggest that $\mathcal{CI}$ failures concentrate in few files, and that focusing pre-integration effort on these might save everyone time; our findings also suggest that poor initial quality in a pull request might be a helpful predictor of eventual quality of the code originating therefrom.

## 2. BACKGROUND AND RELATED WORK

Our work examines the $\mathcal{CI}$ process of handling pull requests ($\mathcal{PR}$s), and its impacts on code quality. The process (Figure 1) begins ① with a contributor forking a GITHUB repository. The contributor makes some changes, and opens a $\mathcal{PR}$ ②, expressing her readiness to have the branch containing her changes merged into the main repository. A $\mathcal{CI}$ system like TRAVIS-CI is then automatically triggered ③. TRAVIS-CI creates a temporary clone of the repository and attempts to merge the $\mathcal{PR}$, then build the system, and run the tests. The results of this process are sent to the core developers ④, who perform manual code review and potentially request more changes ⑤. If core developers decide to accept the $\mathcal{PR}$, typically only if the $\mathcal{CI}$ results are acceptable, it is merged into the main repository ⑥, commonly in the *master* branch. If the $\mathcal{PR}$ code is merged with defects that "escaped" $\mathcal{CI}$ testing, it may eventually trigger bug reports and fixes ⑦; using the version control history and methods like the SZZ algorithm [39], the fix can be traced back to the $\mathcal{PR}$ that inserted the buggy code ⑧.

### 2.1 Continuous Integration and Pull Requests

$\mathcal{CI}$ gained popularity as a component of the Agile movement, in reaction to waterfall-based approaches [42]. Agile rejected the traditional approach to testing at the end of development, after coding. Their goal was to deliver more improvements to customers, in shorter bursts ("sprints") at a faster rate. However, with greater speed comes greater risk of errors. To mitigate this threat, Agile developers spurred the use of frequent, automated testing.

More recently, pull-based development provided another incentive for using $\mathcal{CI}$ & testing: to automate building and testing, so that *more code can be accepted from more people*—a pragmatic, democratizing imperative in open-source projects. On GITHUB, anyone can submit $\mathcal{PR}$s to any project! $\mathcal{CI}$-enabled $\mathcal{PR}$s are becoming mainstream [15, 17, 31, 49].

The value proposition for a GITHUB project to adopt $\mathcal{CI}$ is a bit different than in commercial settings, however. On GITHUB there are two players, core developers and $\mathcal{PR}$ submitters, who experience $\mathcal{CI}$ differently. Core developers can access cloud-based computing resources for testing quite easily; however, constructing a testing harness that ensures rigorous integration, system, and regression testing requires considerable effort. This is needed sustain quality with the rapid pace of current development. Another benefit is the early, automated screening of $\mathcal{PR}$ submissions, so that time is not wasted on poor submissions. With the accelerating usage of $\mathcal{PR}$s [15], $\mathcal{CI}$ is perceived as a mechanism to ensure effectiveness in processing and quality control [31,49]. We also found positive evidence for productivity and quality effects associated with usage of $\mathcal{CI}$ in GITHUB [46]: project teams using TRAVIS-CI integrate more $\mathcal{PR}$s per unit time; users do not report an increasing number of defects as projects adopt $\mathcal{CI}$ and accelerate the rate at which they accept $\mathcal{PR}$s.

For $\mathcal{PR}$ submitters, $\mathcal{CI}$ provides the benefit of quick, extensive feedback from a standardized, rigorous testing regime, that would be more difficult for each submitter to replicate on their own. Still, they must also build tests along with their code, given the common expectation (project-specific) that code must be accompanied by tests. Furthermore, the anticipation of a rigorous $\mathcal{CI}$ evaluation, and presumably the fear of losing face, might incur greater effort at desk-checking (and developing & running unit tests) prior to submission.

Our research here also relates to *testing effectiveness*. The goal of testing is to find faults, before customers find them. Ways to improve testing are of abiding interest, and there is a rich literature on various types of white-box and black-box testing. *Evaluating* these methods is a matter of determining how *effective* (in time, cost, and thoroughness) they are in finding faults before the users experience the related failures. Malaiya [21], starting in 1994, was among the earliest to develop well-founded, empirically validated models of the relationship between code-coverage (white-box) testing and defect finding. There is considerable follow-on work in this vein, that examines connections between code-coverage during testing and eventual quality; *e.g.*, Nagappan *et al* [28] report that STREW-J, a static collection of metrics that measures test case and source code size and complexity, can successfully predict the total number of field failures (eventual quality), and even which components would be more reliable. Other studies of the effectiveness of testing procedures exist but, in general, many of these were constrained by very limited availability of *test results* data, for tests run during development. In general, only results on a few proprietary commercial projects were available for analysis.

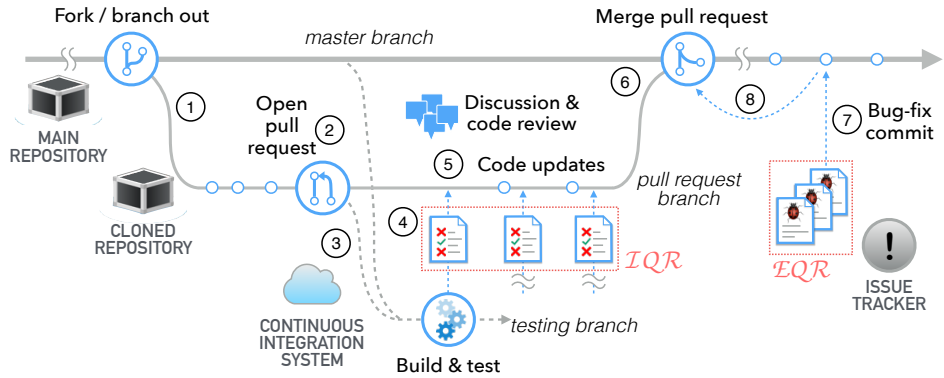

Figure 1: Overview of a $\mathcal{CI}$-enabled $\mathcal{PR}$ process.

With $\mathcal{CI}$ in GITHUB, we have tens of thousands of $\mathcal{PR}$s, each with data on *initial quality* and *eventual quality*. We note here our notion of $\mathcal{IQR}$ comprises failures arising from both build and test steps; we estimate (from informal sampling) that about 45% of the failed $\mathcal{PR}$s include test failures. Thus our data provides an opportunity to shed light on the effectiveness of testing in the GITHUB $\mathcal{PR}$ model, on a much larger scale than before. Large sample sizes, and the diversity of projects in GITHUB, allow us to control for many confounds, and to understand the factors that influence initial quality, and their influence on eventual quality.

## 2.2 Development of Research Questions

The *initial quality* of a $\mathcal{PR}$ is recorded in the pre-merge defect record ($\mathcal{IQR}$), as revealed by the $\mathcal{CI}$ automated build & test procedure. Once merged, the $\mathcal{PR}$ code's eventual quality ($\mathcal{EQR}$) can be found using subsequent defect repairs, using blaming methods like SZZ. Having such a detailed picture of defects and defect fixing code, attributed at line level, gives us an opportunity to study and finely dissect the benefits of $\mathcal{CI}$ within the $\mathcal{PR}$ process.

Several studies [9, 19, 29] revealed that the overall distribution of post-release faults is *highly* skewed, with a few files infested with the bulk of the defects. In contrast, not much information is available about the distribution of build and integration-time failures. The $\mathcal{IQR}$ provides one view into such development-time errors. Understanding these distributions can be an effective way of directing quality-control efforts. If $\mathcal{CI}$ failures also occur in *"hotspots"*, this might help core developers improve system design and/or documentation; and it might help alert $\mathcal{PR}$ submitters when they are touching a file that is rife with previous $\mathcal{CI}$ failures.

*RQ1. How does the distribution of $\mathcal{CI}$ failures vary across source files?*

Once a $\mathcal{PR}$ has been tested by the $\mathcal{CI}$ regime, and perhaps subject to code review, it could be merged into the main repository (we only consider merges into the *master* branch here). After merging, remaining defects within the code added by the $\mathcal{PR}$ may still be discovered, and subsequently fixed. The extent of these repairs reflect the eventual quality of the system, and constitute the $\mathcal{EQR}$.

A good $\mathcal{CI}$ process should achieve high eventual quality. Finding as many defects as possible early would result in clean code entering the main branch, which would not require (m)any fixes. We seek to understand how the $\mathcal{CI}$ results ($\mathcal{IQR}$) affect eventual quality ($\mathcal{EQR}$). Using GIT features, it is possible to trace back from bug-fixes to the $\mathcal{PR}$s from whence the buggy lines originated. This allows us to determine associations between properties of the $\mathcal{PR}$ and the $\mathcal{IQR}$, on the one hand, and the $\mathcal{EQR}$, on the other.

Naturally, many factors may contribute to the eventual quality. Given a specific $\mathcal{PR}$, do previous $\mathcal{CI}$ failures associated with the files affected by a $\mathcal{PR}$ play a role? Does the number of previous authors to these files? Etc. We control for a range of phenomena already known to affect quality within a multi-variate regression model, relying on large sample sizes and available variances to manage the degrees of freedom, and attempt to determine the strength of association between initial quality and eventual quality.

*RQ2. How does the initial quality ($\mathcal{IQR}$) of a $\mathcal{PR}$ relate to the eventual quality ($\mathcal{EQR}$) of the $\mathcal{PR}$?*

Finally, we look at how the properties of the $\mathcal{PR}$ relate to its fate under $\mathcal{CI}$, *viz.*, the odds of failure. There is considerable prior research on defect prediction [20, 26, 34, 52]; however, earlier studies have not specifically targeted $\mathcal{CI}$ failures. To our knowledge, we are the first to attempt to understand the factors influencing $\mathcal{CI}$ failures within the $\mathcal{PR}$ context. We therefore consider the usual factors that are known to be effective in defect prediction, such as size, churn, developer experience, social status, previous failures, and so on, to understand the effects thereof. This results in a modeling exercise with a large set of predictor variables, and is an exploratory study; however, the large number of $\mathcal{PR}$s, and the substantial variance in the predictors, would allow us to understand the effects of several different factors, within one multi-variate regression model.

*RQ3. What are the factors that affect the initial quality of a $\mathcal{PR}$? Do classic predictors used for bug prediction have consistent influence?*

## 3. METHODS

We seek to understand the nature of $\mathcal{CI}$ failures in the $\mathcal{IQR}$, and how these failures relate to the eventual, post-merge quality in the $\mathcal{EQR}$. To this end, we performed both quantitative and qualitative analysis: large-scale regression modeling and a qualitative study of 50 pull requests.

### 3.1 Metrics

Roughly speaking, the metrics can be classified into product metrics, process metrics, churn metrics, social metrics, and our novel $\mathcal{CI}$ metrics.

**Product metrics:** Product properties, *e.g.*, size, are known to have a strong bearing on defect occurrence, and can con-found our study; they must be included as controls in our model [8]. In addition to source files, the size of *test files* can also be expected to impact the effectiveness of $\mathcal{CI}$. We collected information on both source code and test-related files. First, starting with the creation date of each project on GITHUB, we collected 1-month snapshots of their source repository. For each snapshot, we identified source and test files using the conventions of folders organization and file name extensions on GITHUB [14,44,51]. In general, we identify files as test files if they are located in the */test/* or */spec/* folders, or their filenames contain the word "test" as prefix or suffix. To find additional test files, we considered well-known automatic build tools, designed to work with TRAVIS-CI, for all the programming languages in our data set (*e.g.*, Grunt[2] for JavaScript). For each build tool, we found the test files under the specific folders where build tools can automatically detect them. The number of executable lines of code in files were measured using $CLOC$[3]. Thus, for each project we gathered the following metrics, in monthly intervals:

- **proj_src_loc**: number of executable lines of code in source files; measures the size of the project.
- **proj_test_loc**: number of executable lines of code in test files; measures the extent/detail of existing test suites.

***Process metrics***: Our process metrics capture the history of development activities related to a $\mathcal{PR}$. Prior work on defect prediction modeling suggests that process metrics are strongly related to defect occurrence [33]. We collected:

- **n_past_contrs**: total number of contributors to files affected by the $\mathcal{PR}$. It is known that the number of contributors can affect quality [4].
- **pr_hotness**: number of commits to files affected by the $\mathcal{PR}$ during the last month. The frequency of changes is known to affect quality [26].
- **n_old_bugs**: number of eventual bugs in files, which have been fixed before the $\mathcal{PR}$ creation. Defects tend to occur in files where they occurred before.

In general, histories (such as changes, authors, bug-fixes *etc*) are aggregated on a per file basis (*e.g.*, number of commits touching a file). However, since the relation between $\mathcal{PR}$s and source files is one-to-many (*e.g.*, one $\mathcal{PR}$ can touch many files), we need to map the file-level metrics to $\mathcal{PR}$ level. *E.g.*, for *n_past_contrs*, for each file, we use the change log (from *git log*) to determine all contributors who submitted at least one commit to the master branch before the $\mathcal{PR}$ creation. We count contributors only once, even though they may have touched several files associated to the $\mathcal{PR}$. We use the same approach to aggregate other process metrics from the associated files to a $\mathcal{PR}$.

***Churn metrics***: We use churn to measure the size and complexity of a $\mathcal{PR}$.

- **churn_additions**: number of lines added by the $\mathcal{PR}$.
- **churn_deletions**: number of lines deleted by the $\mathcal{PR}$.

***Social metrics***: Human factors are known to impact software development, especially on social coding platforms [7, 27]. We considered the following metrics:

- **n_owner_commits**: total number of commits made by a $\mathcal{PR}$ submitter, before the $\mathcal{PR}$; measures their expertise.
- **n_owner_followers**: total number of GITHUB developers following the submitter of a $\mathcal{PR}$, before the $\mathcal{PR}$ creation, as a measure of the submitter's reputation.

- **core_team** (*binary*): indicates if the submitter was a "core team" member; to this end we check if she had write access to the code repository, or had closed issues and $\mathcal{PR}$s submitted by others, before the $\mathcal{PR}$ creation.
- **n_comments**: number of comments before the $\mathcal{PR}$ was closed. This factor is only used in the eventual quality model because some reviewers comment on the $\mathcal{PR}$ only after receiving the $\mathcal{CI}$ testing results.

*The above are all controls*; next are the main *experimental* variables of interest to this study.

$\mathcal{CI}$ **metrics**: The longer a project uses $\mathcal{CI}$, we can the more maturity we can expect of the $\mathcal{CI}$ process; and thus we expect growing effectiveness. We measure:

- **ci_age**: the time from the project adoption of $\mathcal{CI}$ to the $\mathcal{PR}$ creation, in months.
- **ci_config_lines**: number of lines in the $\mathcal{CI}$ configuration file. TRAVIS-CI is configurable through the *.travis.yml*[4] file. The more TRAVIS-CI is customized, the more extensive, detailed and uncoupled the build and test procedure are. We used the number of customized lines in a $\mathcal{CI}$ config file as an approximate measure of the degree of customization and matureness of the $\mathcal{CI}$ environment.
- **n_old_failures**: number of past TRAVIS-CI failure records for files, before the $\mathcal{PR}$ creation. Previous bugs are known to be an excellent predictor of future bugs [19]; would prior $\mathcal{CI}$ failures predict future $\mathcal{CI}$ failures?
- **ci_running_time**: the duration of the $\mathcal{CI}$ run, in minutes. If the $\mathcal{PR}$ is run through $\mathcal{CI}$ many times, we use the longest duration. The TRAVIS-CI running time is used an an indirect measure of the rigor of the $\mathcal{CI}$ evaluation; longer $\mathcal{CI}$ running times suggests that more test cases are associated with the $\mathcal{PR}$.

### 3.2 Pre-merge Defect Record

For a $\mathcal{PR}$, the $\mathcal{CI}$ results can reveal its initial quality. There are three types of TRAVIS-CI outcomes: *error*, *passed*, and *failed*. If any problem arises before the $\mathcal{CI}$ scripts (including build and test) are run, *e.g.*, a branch has been already deleted or config files are missing, an *error* is returned. Otherwise, TRAVIS-CI builds the project, runs the test cases, and then returns a final result *passed* or *failed*. Project managers can optionally define a $\mathcal{CI}$-run policy where testing is allowed to fail without causing the entire process to fail. If the submitter updates the $\mathcal{PR}$ by pushing new commits, or if the project managers restart the testing process, TRAVIS-CI will return new $\mathcal{CI}$ results. Thus, one $\mathcal{PR}$ could have multiple $\mathcal{CI}$ results. We cannot factually obtain the $\mathcal{PR}$'s $\mathcal{IQR}$ when TRAVIS-CI returns an *error*, because the build and test procedure has not been run. Therefore, we omitted $\mathcal{PR}$s which only contain *error* records; in other words, *we ignore all $\mathcal{PR}$s that $\mathcal{CI}$ failed to start checking over*. In this paper, our $\mathcal{IQR}$ contains both build failures and test failures exposed by TRAVIS-CI. From our experience, most of failures in our data set, especially for dynamically typed languages, belongs to test failure (see our discussion in Section 4.4). A study focusing on build failures in GITHUB is left for future work, and it is interesting to compared the results with the related study of build errors at Google [38].

In theory, an effective $\mathcal{CI}$ process would require that developers fix the problems in a $\mathcal{PR}$ before merging it; defective $\mathcal{PR}$s should be rejected to sustain the code base's quality. In practice, we found that 8% of $\mathcal{PR}$s in our data set failed $\mathcal{CI}$, yet were subsequently merged without being fixed; effectively, they were treated *as if they actually passed*[5]. To distinguish the above kinds of $\mathcal{PR}$s, we collected the **ci_result** variable, which takes one of three values: PASSED, if the $\mathcal{PR}$ passed $\mathcal{CI}$; SOFT FAILURE, if the $\mathcal{PR}$ failed $\mathcal{CI}$, but was merged without fixing; and HARD FAILURE, if the $\mathcal{PR}$ failed $\mathcal{CI}$, and then was fixed before merging.

---
[2] http://gruntjs.com/
[3] http://cloc.sourceforge.net/
[4] http://docs.travis-ci.com/user/customizing-the-build/

## 3.3 Post-merge Defect Record

We adapted the original SZZ algorithm [40] to identify the buggy $\mathcal{PR}$s which are blamed for the $\mathcal{EQR}$. In the first phase, we detected bug-fixing commits and linked them with corresponding issues in the GITHUB issue tracker. We began this phase by extracting the numerical *issue_id*, by searching for specific textual patterns [25,32,40] in project commit logs (*e.g.*, #*issue_id*, fix *issue_id*, close *issue_id*, etc). Then, we cross-checked with the issue tracker to confirm whether such an issue exists and whether it is a bug issue. To mitigate the risk of bias arising from missing and incorrect links [1,3,35] we manually augmented linking rates. We reviewed commits and issues in projects with low linking rates to improve our textual patterns, and found some additional, specific ways to automatically reference a GITHUB issue or close it in a *git* commit (*e.g.*, *gh-issue_id*[6]). Finally, we reconfirmed and updated our textual patterns via random sampling, and removed spurious linking as much as possible.

In the second phase, we identified buggy $\mathcal{PR}$s and then used GIT to identify fix-inducing code. By using *git diff*, we found change details used to fix the linked bug, *i.e.*, files and lines changed in fixing commits. Then, we removed the fixing commits which have not changed any source files (*e.g.*, only changed doc or test files). We used tags to classify issues; however, mis-tagged issues could engender bias [16] in our data. We mitigate this risk by ensuring that: 1) the issue has been tagged as a bug; 2) it was linked to a fixing commit; and 3) developers did indeed change source files to fix it. Next, we used *git blame* to identify the original commits (buggy commits) where the fix-inducing lines had been last added or modified. On GITHUB, the fix-inducing lines might originate either from a direct *git push* or a $\mathcal{PR}$, since developers can contribute code both ways [13]. In this study, we only considered code which came from $\mathcal{PR}$s. For each commit, we identified the originating $\mathcal{PR}$, and the time it was merged. The two timestamps, (the $\mathcal{PR}$ merging time and the corresponding issue open time) are used within the SZZ model to decide if a buggy commit was implicated. We recorded the buggy $\mathcal{PR}$s which contain at least one implicated buggy commit. Finally, we compute the bug-latency between the $\mathcal{PR}$ and the bugs linked to it (issue creation time minus $\mathcal{PR}$ merging time).

Apart from the fact that commits arise from $\mathcal{PR}$s, our approach is the standard SZZ model. We illustrate the process with the example in Figure 2. Following our first phase, let commit *n* be linked with bug *#521*. Then, we identified three fix-inducing lines from the *diff* between commits *n* and *n-1* (*lines-80* and *86* in *file1*, *line-12* in *file2*). Next, by blaming the old changes, we found that (*file1, line-80*) was associated with commit *i* in $PR_1$, and the other two fix-inducing lines—with commits *b* and *g* in $PR_2$. Even though commit *i* finished before issue *#521* was reported, we considered $PR_1$ as not linked to this bug because $PR_1$ had been merged (at *k*) after issue *#521* was reported. Similarly, we considered $PR_2$ the buggy $\mathcal{PR}$ associated with issue *#521*.

---

[5]This suggests that core teams view some $\mathcal{PR}$'s as valuable enough to include immediately in the main branch, perhaps then continuing to work on them later. We found that some failed yet merged $\mathcal{PR}$s are immediately opened as new $\mathcal{PR}$s.
[6]Also reported here: http://goo.gl/QBnk6b

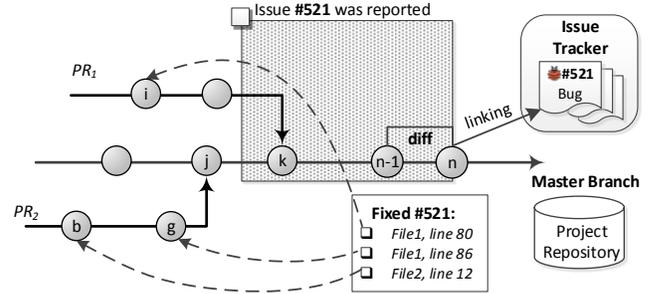

Figure 2: Finding buggy $\mathcal{PR}$s.

Using this methodology, we recorded the eventual quality of a $\mathcal{PR}$ as a binary variable **is_buggy**.

## 3.4 Data Set

In prior work [46, 49, 50], we created a large $\mathcal{PR}$ data set of 246 popular projects in GITHUB, for each including the history from their inception until October 11, 2014. All of them use the pull-based development model, and have $\mathcal{CI}$ enabled using TRAVIS-CI, the most popular $\mathcal{CI}$ server in GITHUB [45, 49]. In this study, our extra requirements, derived from the process of identifying post-merged defect, were to identify projects that: 1) have a sufficient number of issues hosted on GITHUB's issue tracker; 2) the issues are type-tagged at a rate high enough to identify bug issues, and distinguish them from other issue types (*e.g.*, refactor, feature requests, enhancement, etc.); and 3) there is a sufficient number of *resolved* bug issues that have been fixed.

Starting from our existing data set, we chose projects having over 100 issue reports in the issue tracker, with over 50% of the issues tagged with their types.

However, different projects use different (synonymous) labels for the same kind of issue. To resolve this, we manually compiled a list of bug-related keywords as per [46] (*defect, error, bug, issue, mistake, incorrect, fault,* and *flaw*), proceeded by lowercasing and stemming. We identified tags matching these stems in all projects and tagged an issue as bug if at least one stem matched. We found that 27% of issue tags contained at least one bug-related keyword.

Finally, we linked bug issues with bug-fix commits by checking if the issue-id is mentioned in the commit log [10] (details above in Section 3.3), and then filtered out the projects with low linking rate to mitigate the risk of sampling biases (*e.g.*, bug feature and commit feature bias [3]): we required that at least 23.3% of the bugs (the top quartile of our data set) were linked to at least one bug-fix commit.

This selection process left us with 87 projects that utilized $\mathcal{PR}$s and TRAVIS-CI; managed most issues with tags in GITHUB; and had a desirable linking rate between bug-fix commits and bug reports. Overall, we collected 44,193 raw $\mathcal{PR}$s from those projects, and gathered the following data for each $\mathcal{PR}$: initial quality record (*ci_result* in Section 3.2), eventual quality record (*is_buggy* along with bug-latency in Section 3.3) and other metrics described in Section 3.1. The preprocessing steps for initial quality model (Table 4) and eventual quality model (Table 3) are described as follows.

Table 1: Summary statistics for various metrics related to the $\mathcal{IQR}$ model (87 projects; 21,119 rows; a row is a $\mathcal{PR}$).

| Statistic | Mean | St. Dev. | Min | Median | Max |
|---|---|---|---|---|---|
| proj_src_loc | 132,280.5 | 174,938.6 | 510 | 64,419 | 763,973 |
| proj_test_loc | 26,563.3 | 39,519.3 | 0 | 12,734 | 375,642 |
| pr_hotness | 12.50 | 26.23 | 0 | 5 | 286 |
| n_past_contrs | 10.44 | 12.41 | 0 | 6 | 165 |
| n_old_bugs | 4.50 | 15.81 | 0 | 0 | 292 |
| churn_additions | 132.75 | 285.07 | 1 | 31 | 2,884 |
| churn_deletions | 57.99 | 164.97 | 0 | 7 | 1,700 |
| n_owner_commits | 1,510.70 | 2,206.05 | 0 | 762 | 18,448 |
| n_followers | 47.19 | 140.32 | 0 | 12 | 2,221 |
| core_team | 0.65 | 0.48 | 0 | 1 | 1 |
| n_comments | 3.39 | 6.39 | 0 | 1 | 134 |
| ci_age | 12.04 | 8.13 | 1 | 11 | 39 |
| ci_config_lines | 14.51 | 12.04 | 0 | 11 | 108 |
| ci_running_time | 12.70 | 12.31 | 0.02 | 9.07 | 64.92 |
| n_old_failures | 22.03 | 49.06 | 0 | 5 | 665 |

Table 2: Summary statistics for various metrics related to the $\mathcal{EQR}$ model (64 projects; 10,932 rows; a row is a $\mathcal{PR}$).

| Statistic | Mean | St. Dev. | Min | Median | Max |
|---|---|---|---|---|---|
| proj_src_loc | 143,020.8 | 195,034.0 | 857 | 58,540 | 763,973 |
| proj_test_loc | 21,483.2 | 33,387.8 | 0 | 11,637 | 372,788 |
| pr_hotness | 11.62 | 18.03 | 0 | 5 | 177 |
| n_past_contrs | 10.20 | 10.65 | 0 | 6 | 100 |
| n_old_bugs | 3.93 | 13.39 | 0 | 0 | 196 |
| churn_additions | 121.69 | 259.14 | 1 | 28 | 2,463 |
| churn_deletions | 48.63 | 134.45 | 0 | 6 | 1,360 |
| n_owner_commits | 1,237.68 | 1,472.28 | 0 | 688.5 | 8,590 |
| n_followers | 36.54 | 56.79 | 0 | 16 | 477 |
| core_team | 0.64 | 0.48 | 0 | 1 | 1 |
| n_comments | 3.00 | 5.04 | 0 | 1 | 59 |
| ci_age | 10.46 | 6.85 | 1 | 10 | 33 |
| ci_config_lines | 12.99 | 10.43 | 0 | 10 | 87 |
| ci_running_time | 10.63 | 10.70 | 0.12 | 7.52 | 61.92 |
| n_old_failure | 14.67 | 29.75 | 0 | 4 | 289 |
| | PASSED | | SOFT FAILURE | | HARD FAILURE |
| ci_result | 8,891(81.33%) | | 760(6.95%) | | 1,281(11.72%) |

First, contributors may submit source code, test code, and documentation via $\mathcal{PR}$s. We only considered $\mathcal{PR}$s that changed at least one source file, and have been merged into the master branch, because these live on, and are more likely to affect project quality for more users. This step left us 21,119 $\mathcal{PR}$s, after statistical outlier removal (see next Section) for initial quality model, as shown in Table 1.

Then, to model the eventual quality of $\mathcal{PR}$s, we only count bugs in the six-month period after the $\mathcal{PR}$s were merged (similar to prior work [22, 23]). Our rationale for this arose from the concern that bugs delayed longer than six months after the $\mathcal{PR}$ merged, may have other, distinct causes than the code in the merged $\mathcal{PR}$. Then, we removed both clean and buggy $\mathcal{PR}$s merged during the very last six months of our data set, to make sure that developers would have had enough time to find bugs on the rest of $\mathcal{PR}$s (and thus mitigate the effect of right censorship). After the above preprocessing, the $\mathcal{EQR}$ data set (summarized in Table 2) contains 10,932 $\mathcal{PR}$s from 64 projects.

### 3.5 Analysis

***Concentration of $\mathcal{CI}$ Failures***: To answer **RQ1**, we collected all source files in each project's latest version, at the time of data collection. Then, for each file in a project, we collected two measures: a count of $\mathcal{CI}$ failure records (as mined from TRAVIS-CI logs), and a count of eventual bugs, as inferred using our adaptation of the SZZ algorithm (we only counted bugs blamed to $\mathcal{PR}$ commits).

We quantify the dispersion in bug concentration and $\mathcal{CI}$-failure concentration for a project using the *Gini coefficient*[7] [12] of the distributions of the two counts over the project's files. We test for differences in medians between the two Gini distributions (one value per project) using the non parametric *Wilcoxon-Mann-Whitney* (WMW) test.

***Regression Models***: For **RQ2** and **RQ3**, we used mixed effects logistic regression because our outcomes (*ci_failed* for **RQ2**, *is_buggy* for **RQ3**) are dichotomous and the data set is grouped by projects (*i.e.*, several $\mathcal{PR}$s arise from the same project): there may be project-level effects unrelated to the predictor variables we use. In our models, the metrics described in Section 3.1 were modeled as fixed effects, and the variable *proj_id* as a random effect. By examining the distribution of the count metrics, we log transformed them to stabilize the variance and improve model fit [6]. We also checked if the pairwise correlations among our metrics are below 0.6, and the variance inflation factors (VIFs) for each predictor are below 3, to avoid multicollinearity problems [6]. We report for each measure its association with the outcome as an odds ratio, standard error, and significance level. The odds ratio can be calculated from the logistic regression model. *Viz.*, if a coefficient is $-0.5$ for a regressor, a unit change in that regressor, all else held constant, corresponds to $e^{-0.5}$ change in the overall odds ratio (odds ratio is multiplied by about 0.606). As all the count-based regressors (*e.g.*, **proj_src_loc**) are log transformed; a log-ed term in our models shows the multiplicative increase in the response for every $e$ factor increase in the predictor, where $e = 2.718$ is the base for the natural logarithm. From the odds ratio, we can tell how much higher or lower are the odds that a $\mathcal{PR}$ fails $\mathcal{CI}$ or is associated with eventual bugs, per "unit" (*i.e.*, $e$ factor) of the measure.

We also present the area under the *ROC* curve (*AUC*) [36] as an assessment of the goodness of fit for the logistic regressions. Our models were fitted using the R *lme4* [2] package.

***Outlier Removal***: In regression analysis, few data points can disproportionately impact the regression line's slope, artificially inflating the model's fit. This can occur when explanatory variables have highly skewed distributions, as most ours do (*e.g.*, very few $\mathcal{PR}$s have an extremely large number of churned lines of code). Whenever a variable, say $x$, was well fitted by an exponential distribution we identified and rejected as outliers, values that exceeded $k(1 + 2/n)median(x) + \theta$ [30], where $\theta$ is the exponential parameter [37], and $k$ is computed such that not more than 1.5% of values are labeled as outliers. This reduced slightly the size of the data sets onto which we built the regression models, but improved the robustness of our models to outliers.

## 4. RESULTS AND DISCUSSION

### 4.1 Distribution of CI Failures

Analyzing the distributions of Gini coefficients for $\mathcal{CI}$ failures per file and eventual bugs per file, we observe that both indicate *extremely* high concentration of defects in all 87 projects (median: 0.867 for $\mathcal{CI}$ failures; 0.864 for eventual bugs). The WMW test shows no significant difference ($p = 0.4$) between the two distributions. This suggests that, similarly to a small fraction of files being associated with most eventual bugs, only a few files are associated with the majority of problems encountered during $\mathcal{CI}$. For example, in *ipython*[8] (639 source files), one of the projects in our data set, the "buggiest" 5% of the source files account for 50% of the $\mathcal{CI}$ failures and, similarly, 5% of the source files around 50% of the eventual bugs. This heavy concentration of $\mathcal{CI}$ failures suggests that a few files cause most of the trouble, and indicate that targeted documentation, or re-design efforts might have high payoff for $\mathcal{IQR}$, in terms of saving $\mathcal{PR}$ submitters a great deal of effort.

---
[7]The Gini coefficient is an econometrics measure, used to measure income concentration; the highest Gini coefficient for a country is around 0.6, signifying that a few people have cornered most of the income. See the Wikipedia entry.

Table 3: Models for eventual quality.

| Dependent variable: | is_buggy = TRUE | |
| --- | --- | --- |
|  | Control | Full |
| log(proj_src_loc) | 0.753 (0.079)*** | 0.764 (0.081)*** |
| log(proj_test_loc + 0.5) | 0.999 (0.058) | 1.015 (0.061) |
| log(pr_hotness + 0.5) | 1.268 (0.047)*** | 1.245 (0.051)*** |
| log(n_past_contrs + 0.5) | 0.840 (0.080)** | 0.835 (0.089)** |
| log(n_old_bugs + 0.5) | 1.260 (0.046)*** | 1.273 (0.048)*** |
| log(churn_additions) | 1.479 (0.034)*** | 1.461 (0.034)*** |
| log(churn_deletions + 0.5) | 1.140 (0.028)*** | 1.141 (0.028)*** |
| log(n_owner_commit + 0.5) | 0.904 (0.037)*** | 0.913 (0.038)** |
| log(n_followers + 0.5) | 1.148 (0.041)*** | 1.146 (0.042)*** |
| as.logical(core_team) | 1.163 (0.125) | 1.162 (0.126) |
| log(n_comments + 0.5) | 1.243 (0.041)*** | 1.223 (0.042)*** |
| log(ci_age) |  | 0.902 (0.078) |
| log(ci_config_lines + 0.5) |  | 1.051 (0.097) |
| log(ci_running_time) |  | 0.999 (0.061) |
| log(n_old_failures + 0.5) |  | 1.010 (0.044) |
| **ci_result=Soft Failure** |  | 1.144 (0.167) |
| **ci_result=Hard Failure** |  | 1.315 (0.111)** |
| Constant | 0.173 (0.796)** | 0.136 (0.832)** |
| Observations | 10,932 | 10,932 |
| Area Under the ROC Curve | 87.22% | 87.30% |
| *Note:* | *p<0.1; **p<0.05; ***p<0.01 | |

We further asked if both kinds of failures are concentrated in *the same files*. For each of our 87 projects, we calculated the Spearman rank correlation [41] between initial quality and eventual quality failures for each file in that project. The values of Spearman's $\rho$ range from -0.001 to 0.859 (mean: 0.387; median: 0.386). In only 19 projects the source files have a relatively strong relationship between initial quality and eventual quality (Spearman's $\rho \geq 0.6$ [43]).

> **Result 1**: $\mathcal{CI}$ failures appear to be concentrated in a few files, just like normal bugs are. At file level, $\mathcal{CI}$ failures are not highly correlated with eventual bugs.

Note that this correlation is not strong **at the file level**. Below we explore whether failures in the $\mathcal{IQR}$ are associated with failures in the $\mathcal{EQR}$, *but at the $\mathcal{PR}$ level* instead.

## 4.2 Eventual Quality Model

To answer **RQ2**, we use mixed effects logistic regression to study the relationship of the $\mathcal{IQR}$ of a $\mathcal{PR}$ to its $\mathcal{EQR}$. We build two models, the first with just controls, and the second, a full model, will our $\mathcal{CI}$ related variables (Table 3).

Overall, the two $AUC$s of the eventual quality models indicate a good level of fit, by accepted standards (AUC>80%) [24]. Comparing the full model to the control, we can see that the $\mathcal{CI}$ related variables have limited effect on the prediction of eventual bugs (the $AUC$s for the two models are virtually identical, 87.22% vs. 87.30%). Only the initial quality (**ci_result**) is statistically significant out of the 5 $\mathcal{CI}$-related metrics. The outcome of SOFT FAILURE, when integrators merge failed $\mathcal{PR}$s without first fixing them, is not significantly different from the baseline of a passing $\mathcal{PR}$ (PASSED). However, if the $\mathcal{CI}$ outcome was HARD FAILURE, *i.e.*, the broken build has been fixed by developers, the $\mathcal{PR}$ is 31.5% more likely to have $\mathcal{EQR}$ failures (odds factor 1.315). Since only about 12% of $\mathcal{PR}$s experience HARD FAILURE, this variable's effect, although statistically significant at the 0.05 level, has only a modest effect on overall explanatory power beyond the influence of the controls.

[8]Both Gini coefficients for *ipython* are well below the population's median: $\mathcal{CI}$ failures, 0.797; eventual bugs, 0.792.

To shed more light on the above, we randomly chose 10 $\mathcal{PR}$s with SOFT FAILUREs from our data set (soft failure evidence can be found by checking the $\mathcal{CI}$ logs or the content of code reviews on GITHUB). We found that the failures in these 10 $\mathcal{PR}$s were *extrinsic*, *e.g.*, due to bugs in another module, broken test suites, and service crashes, rather than errors *intrinsic* to the code in the $\mathcal{PR}$. Example comments on these $\mathcal{PR}$s from the core team were:

**Example_1:** "@... Ok this looks fine, the only failures look like that issue you've been looking into for the get_current_page_url issue right?"
"@... Yep, which is an open issue still."
**Example_2:** "Tests timed out on Travis for some reason, but they're passing on my machine. I'm merging this, then."

It is notable that **n_old_failures** is not significant in our model. Thus, although a $\mathcal{PR}$ touched some files which historically had $\mathcal{CI}$ failures, it doesn't affect eventual quality. This is consistent with the result in Section 4.1 that $\mathcal{IQR}$ is not very highly correlated with $\mathcal{EQR}$ at file level.

We turn to the non-$\mathcal{CI}$ (control) metrics in the full model. Among product metrics, the project size (**proj_src_loc**) has a negative association with $\mathcal{EQR}$ failures, *i.e.*, the odds of merged $\mathcal{PR}$s (tested by $\mathcal{CI}$ first) in larger projects having eventual bugs are 0.764 times smaller for every $e$ factor increase in LOC. This may have several explanations. First, contributors in large projects could be more experienced. Second, large projects may have higher code standards for contributors. If $\mathcal{PR}$s cannot meet the projects' requirements, they would be immediately rejected by integrators [15]. Third, large, well-established projects may have better collections of formal on-line documentation, and informal help from archived on-line discussions (accessible via search-engines) that better supports $\mathcal{PR}$ contributors.

As expected, all process metrics are significant. If $\mathcal{PR}$s touch hot files (**pr_hotness**), it is more likely they will have eventual bugs. Having previous bugs is also a strong predictor of a $\mathcal{PR}$'s bugginess; the odds of a buggy $\mathcal{PR}$ increases 1.25 times for each $e$ factor increase in **n_old_bugs**.

Unsurprisingly for the churn metrics, the larger the churn (**churn_additions** and **churn_deletions**) of a $\mathcal{PR}$, the more likely it is to be buggy.

Lastly, three of the social metrics are significant. In general, experienced developers (higher **n_owner_commits**) have lower odds to introduce bugs in $\mathcal{EQR}$. However, the more **followers** a developer has, the higher the odds she will submit a buggy $\mathcal{PR}$. This might be because the influential developers typically undertake difficult, more critical tasks in a given project, which may result in more bugs than on average. Then, from the perspective of eventual quality, there is no significant difference between $\mathcal{PR}$s submitted by **core** developers and those submitted by external contributors, despite the fact that core team $\mathcal{PR}$s are processed quicker and merged preferentially [44, 49]. Our result suggests that the standard quality-control process introduced by $\mathcal{CI}$ ensures uniform quality for both core and external contributors.

Table 4: Models for initial quality.

| Dependent variable: | ci_result = Hard Failure | |
|---|---|---|
| | Control | Full |
| log(proj_src_loc + 0.5) | 0.870 (0.063)** | 0.861 (0.063)** |
| log(proj_test_loc + 0.5) | 1.322 (0.043)*** | 1.150 (0.042)*** |
| log(pr_hotness + 0.5) | 0.926 (0.025)*** | 0.888 (0.027)*** |
| log(n_past_contrs + 0.5) | 1.093 (0.031)*** | 0.973 (0.032) |
| log(n_old_bugs + 0.5) | 1.063 (0.024)*** | 1.024 (0.025) |
| log(churn_additions + 0.5) | 1.679 (0.017)*** | 1.616 (0.018)*** |
| log(churn_deletions + 0.5) | 1.004 (0.014) | 1.004 (0.015) |
| log(n_owner_commits + 0.5) | 0.993 (0.018) | 0.964 (0.019)* |
| log(n_followers + 0.5) | 0.993 (0.020) | 1.019 (0.020) |
| as.logical(core_team) | 0.731 (0.061)*** | 0.729 (0.062)*** |
| log(ci_age) | | 0.849 (0.047)*** |
| log(ci_config_lines + 0.5) | | 0.745 (0.076)*** |
| log(n_old_failures + 0.5) | | 1.175 (0.021)*** |
| log(ci_running_time) | | 2.162 (0.042)*** |
| Constant | 0.007 (0.626)*** | 0.021 (0.634)*** |
| Observations | 21,119 | 21,119 |
| Area Under the ROC Curve | 81.04% | 82.45% |
| *Note:* | *p<0.1; **p<0.05; ***p<0.01 | |

Prior work [7] found that $\mathcal{PR}$s with lots of comments tend to signal controversy on a social coding platform, and also are less likely to be accepted [44] in GitHub. The implication is that the quality of those $\mathcal{PR}$s may be lower. Here, we find that an $e$ factor increase in **n_comments** for a $\mathcal{PR}$ is associated with an increase of its eventual bugginess likelihood by 22.3%.

> **Result 2**: *While $\mathcal{CI}$ metrics in general have a small effect on improving models of eventual bugginess, the odds are 31.5% higher that a $\mathcal{PR}$ will have eventual bugs if it has failed $\mathcal{CI}$. Moreover, product, process, churn, and social metrics have mostly expected effects except, notably, that larger projects reduce the threat on quality of merged $\mathcal{PR}$s.*

It should be noted that while Hard Failure status of the $\mathcal{PR}$ is associated with a 31.5% increase in odds of failure in the $\mathcal{EQR}$, there are overall not many $\mathcal{EQR}$ failures; so the change in the AUC from introducing these $\mathcal{IQR}$ related variables is fairly small.

### 4.3 Initial Quality Model

Finally, we study initial quality ($\mathcal{IQR}$). Are the factors influencing $\mathcal{IQR}$ different from those influencing $\mathcal{EQR}$?

We argue that developers are more concerned about "hard" $\mathcal{CI}$ failures, since they would indeed fix those failures before merging $\mathcal{PR}$s. In addition, we did find above that "hard" $\mathcal{CI}$ failures are associated with eventual failures (Table 3). Therefore, we focus the following discussion only on $\mathcal{PR}$s with $\mathcal{CI}$ outcomes Passed and Hard Failure.

Similarly, we built two models, with and without $\mathcal{CI}$ metrics, to answer **RQ3**. Overall, the two $AUC$s of the initial quality models in Table 4 are over 81%, indicating good model fit. Compared to eventual quality, the model achieves a moderate improvement by including the $\mathcal{CI}$ metrics (81.04% vs. 82.45% in $AUC$). Moreover, all $\mathcal{CI}$ metrics have statistically significant effects.

We note that the dominant factor is **ci_running_time**. A longer $\mathcal{CI}$ running time may indicate longer setup and compilation, or a larger test suite, likely to uncover more faults. Also, as expected, an increase in **ci_age** and **ci_config_lines** is associated with lower likelihood of having $\mathcal{CI}$ failures; the $\mathcal{IQR}$ of $\mathcal{PR}$s should improve with increased $\mathcal{CI}$ usage and the customization degree of the $\mathcal{CI}$ environment.

Interestingly, the higher the number of past $\mathcal{CI}$ failures, **n_old_failures**, associated with a file that the $\mathcal{PR}$ touches, the higher the odds that $\mathcal{CI}$ will fail on this $\mathcal{PR}$ (the odds of failure increase 1.175 times for each $e$ factor increase in past failures), consistent with our findings in **RQ1**. This interesting result is analogous to similar findings for defects in general [19], as well as potentially actionable: core developers could focus attention on these files, to understand and remedy the source of repeated $\mathcal{CI}$ failure.

Two process metrics (**n_past_contrs** and **n_old_bugs**), powerful in bug prediction and significant in our $\mathcal{EQR}$ model, don't have significant effects here on $\mathcal{CI}$ failures. In addition, the $\mathcal{CI}$ failure-prone files aren't necessarily the ones changing a lot. We note that $\mathcal{PR}$s changing hot files (**pr_hotness**) have better odds for initial quality. However, they have much higher odds of being associated with eventual bugs (as shown in the eventual quality model, Table 3). The incongruence between the effect of process metrics on eventual quality and initial quality is consistent with $\mathcal{IQR}$ and $\mathcal{EQR}$ not having much overlap, after $\mathcal{CI}$ adoption, as suggested in **RQ1** also.

Among product metrics, project size (**proj_src_loc**) has a negative effect on $\mathcal{PR}$ failures, associated with a decrease of the $\mathcal{CI}$ failure odds by a factor of 0.861 for every $e$ factor increase in project size. The size of the test suite (**proj_test_loc**), as expected, has strong positive effects on $\mathcal{CI}$ failures. The odds of $\mathcal{CI}$ failure would increase by a factor of 1.150 for every $e$ factor increase in the test suite size, all else held constant, indicating that the projects with higher test coverage are more likely to detect the bugs during the $\mathcal{CI}$ stage.

$\mathcal{PR}$ churn metrics have positive effects, as with regular defects [26]: the odds of $\mathcal{CI}$ failure is 1.616 times higher for every $e$ factor increase in **churn_additions**.

Likewise, the more experienced the submitter (higher **n_owner_commits**), the less likely a $\mathcal{PR}$ is to fail $\mathcal{CI}$ testing. Similarly, the initial quality of $\mathcal{PR}$s by core team members is clearly better than that of $\mathcal{PR}$s from outsiders; the odds of $\mathcal{CI}$ failure decreases by 27.1% when the **core_team** becomes TRUE. The final number of review comments can only be collected after a $\mathcal{PR}$ is merged, but the $\mathcal{CI}$ result is produced before that; so we don't consider that metric here.

> **Result 3**: *A mature $\mathcal{CI}$ process (more customized, more tests) is associated with better fault detection during $\mathcal{CI}$. However, some factors used for bug prediction do not predict the pre-merge defects; in particular, the project size and $\mathcal{PR}$ hotness have completely opposite effects on the $\mathcal{IQR}$, than they do in bug prediction.*

### 4.4 Qualitative Analysis

We qualitatively analyzed a sample from our data to better understand what $\mathcal{IQR}$ issues are found by $\mathcal{CI}$, and why some $\mathcal{PR}$s have $\mathcal{EQR}$ issues even after being vetted by $\mathcal{CI}$.

We identified the top-5 projects with the most $\mathcal{PR}$s having Hard Failures, from among popular projects with at least 1,000 stars on GitHub: *ipython*, *pydata* and *scipy* written in Python, *numpy* in C & Python, and *cakephp* in PHP. For each project, we randomly chose 5 clean and 5 buggy $\mathcal{PR}$s from our data set, respectively. All 50 $\mathcal{PR}$s are required to have Hard Failures; for each $\mathcal{PR}$, we check: 1) its $\mathcal{CI}$ failure logs; 2) GitHub comments related to the failure; 3) the diff between the failed code and subsequent passing code, when the defect was fixed. 4) for the $\mathcal{PR}$s with $\mathcal{EQR}$ problems, we also look at bug reports and the bug-fix commit details (as described in Section 3.3). Following this process, we summarize the typical $\mathcal{IQR}$s failures exposed by $\mathcal{CI}$, and discuss reasons why some $\mathcal{EQR}$s failures escaped $\mathcal{CI}$.

For statically typed languages (*e.g.*, C/C++), $\mathcal{CI}$ failures typically arise from missing headers or compiler errors (*e.g.*, name/type mismatch). These defects are more readily detectable by developers, who can first compile a project locally. For dynamically typed languages (*e.g.*, Python), the above "build" failure would only be exposed if the relevant lines are actually executed, which assumes a well-established testing regime. Otherwise, those $\mathcal{IQR}$s would survive $\mathcal{CI}$ diagnosing, and live to harass users as $\mathcal{EQR}$s.

We illustrate typical examples of $\mathcal{CI}$ failures in Table 5. In Example 1, a contributor tried to rename `using_magics` to `using_paste_magics`, changed lines 161, 163, and 353, but forgot to change the invocation on line 731 (displayed under "Fixing code" in Table 5). $\mathcal{CI}$ helps find this mistake. We also noted that $\mathcal{CI}$ is used by projects supporting multiple language versions (*e.g.*, Python 2.x and Python 3.x), by automatically testing the new code in the corresponding execution environments on the cloud. In Example 2, `NoneType` no longer exists after Python 3.0, so the $\mathcal{PR}$ passed testing for Python 2.6 & 2.7, but failed in 3.0 and above.

We also note that $\mathcal{CI}$ exposes defects due to program logic, wrong parameter transfers, and coding style violations, some of which are hard to detect by humans (see, *e.g.*, the dialogue between reviewers in Examples 3 and 4). Checking the code style is an important task [15, 50] when reviewing code in $\mathcal{PR}$s; $\mathcal{CI}$ helps alleviate this burden, *e.g.*, Pylint, a code convention tool for Python, can be embedded into $\mathcal{CI}$ to check for *pep8* violations (Example 5).

Our quantitative results indicate that $\mathcal{EQR}$ failures aren't highly correlated with $\mathcal{IQR}$ failures. From our qualitative study, we find two major reasons. First, if the $\mathcal{PR}$ contributes to a feature enhancement or an entirely new feature, the test suite may not cover it. Example 1 in Table 6 illustrates a $\mathcal{PR}$ that enhanced the performance of `Series`, and $\mathcal{CI}$ found a defect when `m.groups()[0]` returns an empty string rather than a `None`. One month after the $\mathcal{PR}$ was merged, developers reported a $\mathcal{EQR}$ failure: the code crashed for time series, and there was no test suite to cover this case. They subsequently fixed the error and added new test code, such that $\mathcal{CI}$ would cover future changes to this feature. In the examples we analyzed during our case study, we find that insufficient test code is the main reason for $\mathcal{EQR}$ failures, after initially passing $\mathcal{CI}$.

In addition, $\mathcal{EQR}$ failures may occur in a $\mathcal{PR}$ when the dependent packages, project code architectures, or coupling modules are upgraded. Example 2 in Table 6 illustrates a $\mathcal{EQR}$ failure in *numpy*, when the project tried to build with a newer version of Python. We argue that this kind of $\mathcal{EQR}$ failure is really hard to avoid at $\mathcal{CI}$ time.

## 5. THREATS TO VALIDITY

We refer to prior work for threats to mining GIT [5] and GITHUB [18]. The following are additional weaknesses [47].

***Conclusion Validity:*** These threats refer to issues that affect our ability to draw the correct conclusion about relations between the different metrics and the $\mathcal{IQR}$ and $\mathcal{EQR}$ outcomes. We believe conclusion validity threats are lim-ited. First, our metrics are *reliable*; all are objective measures, and can be replicated. Moreover, our two data sets for initial quality and eventual quality contain tens of thousands of $\mathcal{PR}$s across 87 projects, with substantial variance in the predictors, ensuring high *statistical power*. However, because of this diversity, there is a risk that the variation due to individual differences between projects is larger than due to our main predictors (*random heterogeneity*). In our models, we control for variation due to individual differences between projects explicitly, using a random effect. Finally, we checked for compliance with *lme4*'s *assumptions* when building our models: absence of multicollinearity, agreement between QQ-plots and a normal distribution, residuals with no difference in variance variability across the range.

Table 5: Examples of $\mathcal{IQR}$ failures and fixes.

| |
|---|
| **1: PR #2744 in *ipython*. $\mathcal{IQR}$:** Undefined variable error |
| **Travis-CI log:** |
| *[...] object has no attribute 'using_magics'* |
| **Fixing code:** |
| −   if not self.using_magics: |
| +   if not self.using_paste_magics: |
| **2: PR #7081 in *pydata*. $\mathcal{IQR}$:** Undefined class error |
| **Travis-CI log:** only in Python: 3.2 & 3.3 |
| *ImportError: cannot import name NoneType* |
| **Fixing code:** |
| − isinstance(colspec[0], (int, np.integer, NoneType)) |
| + isinstance(colspec[0], (int, np.integer, type(None))) |
| **3: PR #4466 in *numpy*. $\mathcal{IQR}$:** Parameter transfer error |
| **Reviewers dialogue:** |
| COMMENT 1: *[...] It is the source of the test failure, the copied iterator was not getting any values cached.* |
| COMMENT 2: *Good example of why not to copy quickly copy paste stuff. [...] couldn't imagine it causing things going bad.* |
| **Fixing code:** |
| − if (npyiter_cache_values(self) < 0) { |
| + if (npyiter_cache_values(iter) < 0) { |
| **4: PR #5158 in *ipython*. $\mathcal{IQR}$:** Logic error |
| **Reviewers dialogue:** |
| COMMENT 1: *[...] there is a test failure on travis. but this one is weird [...] and probably worth looking into* |
| COMMENT 2: *looks like a real error on my part, fix incoming* |
| **Fixing code:** |
| − if is_hidden(os_path, self.notebook_dir): |
| + if not os.path.isdir(os_path): |
| +   raise web.HTTPError(404, u'dir does not exist: %r'%os_path) |
| + elif is_hidden(os_path, self.notebook_dir): |
| **5: PR #3407 in *scipy*. $\mathcal{IQR}$:** Coding style |
| **Reviewers dialogue:** |
| COMMENT 1: *Can you undo this change? Violates PEP8. Travis pep8 check fails on this one. Remove backslash.* |
| **Fixing code:** |
| − raise AssertionError("First warning for %s is not a " \\     "%s( is %s)" % (func.__name__, warning_class, l[0])) |

***Internal Validity:*** These threats arise when an observed relationship between independent and dependent variables might be an artifact of *confounding variables*. To mitigate, we include controls in our models, to capture product, process, and social aspects that could have confounded the relationship between $\mathcal{CI}$ failures and eventual quality.

Another important source of internal validity threats is *instrumentation*. To mitigate these, we performed extensive code reviews of, and iterative updates to, all our data extraction (Java & Python) and analysis (R) scripts (at least one co-author reviewed each script produced by another), as well as extensive spot-checking of our data, on pilot samples.

We also acknowledge threats related to *maturation*. Our $\mathcal{PR}$ data spans four years, during which the $\mathcal{CI}$ process and configurations could have matured, potentially changing the meaning of our metrics from early history to later.

Table 6: Examples of $\mathcal{EQR}$ failures and fixes.

| **1: PR #5944 in *pydata*. $\mathcal{IQR}$ type: Logic error** |
|---|
| REVIEWER COMMENT: |
| *m.groups()[0] might return ''.* |
| *Only None should be turned into NaN.* |
| $\mathcal{IQR}$ Fixing code: |
| `− return m.groups()[0] or NA` |
| `+ return [np.nan if it is None else it for it in m.groups()]` |
| $\mathcal{EQR}$ **reason:** No test case for a new feature |
| **Bug report #6348:** |
| *series.str.extract does not work for timeseries.* |
| $\mathcal{EQR}$ Fixing code: |
| `result = Series([f(val)[0] for val in arr],` |
| `        name=regex.groupindex.get(1),` |
| `+        index=arr.index)` |
| `+ def check_index(index): ...` |
| **2: PR #3507 & #3517 in *numpy*** |
| $\mathcal{EQR}$ **reason:** Dependency upgrading |
| **Bug report #3598:** *Build fails with Python 3.4a1.* |
| **Discussion of the bug:** |
| COMMENT 1: *Py3.4 doc says "No build and C API changes", so I'd think the compile args shouldn't change. But I can't find so quickly who decided this and why.* |
| COMMENT 2: *It's a Numpy bug, we don't allow that construct.* |
| $\mathcal{EQR}$ **fixing message:** |
| *BUG: Fix variable declaration after statement.* |
| *Some declarations that are not at the beginning of a block have slipped into the code. This breaks compilation on Python3.4a1. The Numpy coding standard also disallows that construct.* |

***Construct Validity:*** Construct validity concerns measurement and whether our metrics actually capture the properties we intend to capture. We acknowledge several threats. First, we used the total number of executable lines of code in test files as a measure of the size and quality of existing test suites. Based on this, we assumed that $\mathcal{PR}$s submitted to projects with larger test suites are more likely to be more thoroughly tested, other things being equal, without linking the contents of the $\mathcal{PR}$ to individual test cases. This assumption, although not uncommon, may not always hold.

Second, our definition of $\mathcal{IQR}$ failures as "broken" TRAVIS-CI builds is quite general: a TRAVIS-CI failure could be either a test failure or a build failure (at least for statically typed languages requiring a separate compilation step); this could have affected the strength of the association between $\mathcal{IQR}$ failures and $\mathcal{EQR}$ failures, since eventual failures are less likely to be due to compilation issues. A more fine-grained analysis is left for future work; here we limit to reporting that, through informal observation, we found that more than 80% of the $\mathcal{PR}$s with TRAVIS-CI failures in a sample from our data contain actual test failures.

Third, we evaluate a $\mathcal{PR}$'s initial quality based on its $\mathcal{CI}$ failure record. However, a $\mathcal{CI}$ failure could also be attributed to a broken or obsolete test, not just a $\mathcal{PR}$ defect. Based on informal observation, we believe these cases to be infrequent.

Fourth, we made a number of choices when aggregating data, *e.g.*, we chose the $\mathcal{CI}$ time as the longest of potentially multiple TRAVIS-CI runs, and we only counted defects in the six-month period after a $\mathcal{PR}$ was merged. We believe our operationalization is reasonable and, whenever available, we based it on previous work. Still, it is unclear whether a different operationalization would affect our conclusions.

Fifth, not all user-found defects are reported in issue trackers; however, we only analyzed popular projects with active issue trackers, where the discovery chance may be higher.

Finally, $\mathcal{PR}$s without $\mathcal{CI}$ failures are not necessarily never-failed $\mathcal{PR}$s. Informally, we came across $\mathcal{PR}$s without $\mathcal{CI}$ failures, but that have been previously submitted, $\mathcal{CI}$-tested, found to be defective, repeatedly discussed, and updated, before being resubmitted again "fresh" (different $\mathcal{PR}$ ID and different commit SHA), after improvements. Such $\mathcal{PR}$s could have affected our eventual quality models, where $\mathcal{CI}$-related predictors were used. Based on our informal data explorations, we believe these occurrences to be rare.

***External Validity:*** Empirical studies which rely on activity traces have a number of known threats, which limit their *generalizability* much beyond the data set examined [48]. Assuming no global bias is present, though, the current results confirm and extend our previous studies on this topic, and combined with them offer a consistent and robust story.

Still, it is worth noting that while canonical $\mathcal{CI}$ assumes that one integrates all the time, *i.e.*, with each commit, in pull-based development one typically opens a $\mathcal{PR}$ only after the development of a specific task is finished (perhaps multiple commits). This "not really continuous" $\mathcal{CI}$ may further limit the generalizability of our findings beyond GITHUB.

## 6. CONCLUSIONS

Software developers have always yearned to produce more code without sacrificing quality. We have in our previous work shown that automating testing and building, via $\mathcal{CI}$ after each pull request, does provide a scalable technology to increase code contributions to the project with no appreciable impact on end user defects, cumulatively.

In this paper we expanded on that work by studying the nature of $\mathcal{CI}$ detected defects, on per pull request basis, *i.e.*, where they occur in code, how many are there, what process, product, and social factors are associated with them, and how they relate to eventual bugs. We found that $\mathcal{CI}$ detected bugs are distributed just like normal bugs, and thus should be no easier to detect in principle. However, they do not correlate highly with eventual bugs, on a per-file basis.

This concentration of $\mathcal{CI}$ failures into a very few files was previously not known, and is likely to be <u>actionable</u>; by focusing attention on these files, projects and developers are likely to gain tangible improvements in $\mathcal{CI}$ productivity.

We found differences between initial defects and eventual quality defects. In particular, finding new $\mathcal{CI}$ failures don't seem to be informed by the presence of old non $\mathcal{CI}$ bugs. However, occurrence of non $\mathcal{CI}$ bugs in a file's history, is associated with increase in odds of $\mathcal{EQR}$ failures, *even in the presence of $\mathcal{CI}$*[9]. In general, this provides an <u>actionable</u> indication that even in the presence of $\mathcal{CI}$, careful review of files where $\mathcal{EQR}$ problems occurred before is sensible. In addition, HARD FAILURE in $\mathcal{CI}$ is associated with a 31.5% increase in risk of eventual failures, also suggesting that (even if repaired) code in such $\mathcal{PR}$s warrants extra scrutiny.

Finally, the interplay between initial and final quality is interesting, as it seems only weakly related. While we show there is no general trend, we do find some $\mathcal{PR}$s with more problems initially have also more defects eventually, but there are other $\mathcal{PR}$s that don't. A more fine grained examination could potentially associate the $\mathcal{PR}$ type with the ability of $\mathcal{CI}$ to capture eventual problems.

---

[9]The model indicates that a factor $e$ increase in old, historically cumulated non-$\mathcal{CI}$ bugs in a $\mathcal{PR}$ is associated with a 27.3% increase in the risk of $\mathcal{EQR}$ failure.

# 7. REFERENCES

[1] A. Bachmann, C. Bird, F. Rahman, P. Devanbu, and A. Bernstein. The missing links: Bugs and bug-fix commits. In *FSE*, pages 97–106. ACM, 2010.

[2] D. M. Bates. lme4: Mixed-effects modeling with R. 2010.

[3] C. Bird, A. Bachmann, E. Aune, J. Duffy, A. Bernstein, V. Filkov, and P. Devanbu. Fair and balanced?: Bias in bug-fix datasets. In *ESEC/FSE*, pages 121–130. ACM, 2009.

[4] C. Bird, N. Nagappan, B. Murphy, H. Gall, and P. Devanbu. Don't touch my code!: examining the effects of ownership on software quality. In *ESEC/FSE*, pages 4–14. ACM, 2011.

[5] C. Bird, P. C. Rigby, E. T. Barr, D. J. Hamilton, D. M. German, and P. Devanbu. The promises and perils of mining git. In *MSR*, pages 1–10. IEEE, 2009.

[6] J. Cohen, P. Cohen, S. G. West, and L. S. Aiken. *Applied multiple regression/correlation analysis for the behavioral sciences*. Routledge, 2013.

[7] L. Dabbish, C. Stuart, J. Tsay, and J. Herbsleb. Social coding in GitHub: Transparency and collaboration in an open software repository. In *CSCW*, pages 1277–1286. ACM, 2012.

[8] K. El Emam, S. Benlarbi, N. Goel, and S. N. Rai. The confounding effect of class size on the validity of object-oriented metrics. *IEEE Transactions on Software Engineering*, 27(7):630–650, 2001.

[9] N. E. Fenton and N. Ohlsson. Quantitative analysis of faults and failures in a complex software system. *Software Engineering, IEEE Transactions on*, 26(8):797–814, 2000.

[10] M. Fischer, M. Pinzger, and H. Gall. Populating a release history database from version control and bug tracking systems. In *ICSM*, pages 23–32. IEEE, 2003.

[11] M. Fowler. Continuous integration, 2006. http://martinfowler.com/articles/continuousIntegration.html.

[12] C. Gini. Variabilitè e mutabilitè. *Studi Econornico-Giuridici della R. Universita de Cagliari*, 1912.

[13] G. Gousios, M. Pinzger, and A. v. Deursen. An exploratory study of the pull-based software development model. In *ICSE*, pages 345–355. ACM, 2014.

[14] G. Gousios and A. Zaidman. A dataset for pull-based development research. In *MSR*, pages 368–371. ACM, 2014.

[15] G. Gousios, A. Zaidman, M.-A. Storey, and A. Van Deursen. Work practices and challenges in pull-based development: The integrator's perspective. In *ICSE*, pages 358–368. IEEE, 2015.

[16] K. Herzig, S. Just, and A. Zeller. It's not a bug, it's a feature: How misclassification impacts bug prediction. In *ICSE*, pages 392–401. IEEE, 2013.

[17] E. Kalliamvakou, D. Damian, K. Blincoe, L. Singer, and D. German. Open source-style collaborative development practices in commercial projects using GitHub. In *ICSE*, pages 574–585. IEEE, 2015.

[18] E. Kalliamvakou, G. Gousios, K. Blincoe, L. Singer, D. M. German, and D. Damian. The promises and perils of mining GitHub. In *MSR*, pages 92–101. ACM, 2014.

[19] S. Kim, T. Zimmermann, E. J. Whitehead Jr, and A. Zeller. Predicting faults from cached history. In *ICSE*, pages 489–498. IEEE, 2007.

[20] P. Knab, M. Pinzger, and A. Bernstein. Predicting defect densities in source code files with decision tree learners. In *MSR*, pages 119–125. ACM, 2006.

[21] Y. K. Malaiya, M. N. Li, J. M. Bieman, and R. Karcich. Software reliability growth with test coverage. *IEEE Transactions on Reliability*, 51(4):420–426, 2002.

[22] S. McIntosh, Y. Kamei, B. Adams, and A. E. Hassan. The impact of code review coverage and code review participation on software quality: A case study of the Qt, VTK, and ITK projects. In *MSR*, pages 192–201. ACM, 2014.

[23] S. McIntosh, Y. Kamei, B. Adams, and A. E. Hassan. An empirical study of the impact of modern code review practices on software quality. *Empirical Software Engineering*, pages 1–44, 2015.

[24] C. Metz. Basic principles of ROC analysis. *Semin Nucl Med.*, 8:283–298, 1978.

[25] A. Mockus and L. G. Votta. Identifying reasons for software changes using historic databases. In *ICSM*, pages 120–130. IEEE, 2000.

[26] N. Nagappan and T. Ball. Use of relative code churn measures to predict system defect density. In *ICSE*, pages 284–292. IEEE, 2005.

[27] N. Nagappan, B. Murphy, and V. Basili. The influence of organizational structure on software quality: An empirical case study. In *ICSE*, pages 521–530. ACM, 2008.

[28] N. Nagappan, L. Williams, M. Vouk, and J. Osborne. Early estimation of software quality using in-process testing metrics: a controlled case study. In *ACM SIGSOFT Software Engineering Notes*, volume 30, pages 1–7. ACM, 2005.

[29] T. J. Ostrand and E. J. Weyuker. The distribution of faults in a large industrial software system. In *ACM SIGSOFT Software Engineering Notes*, volume 27, pages 55–64. ACM, 2002.

[30] J. K. Patel, C. Kapadia, and D. B. Owen. *Handbook of statistical distributions*. M. Dekker, 1976.

[31] R. Pham, L. Singer, O. Liskin, and K. Schneider. Creating a shared understanding of testing culture on a social coding site. In *ICSE*, pages 112–121. IEEE, 2013.

[32] F. Rahman and P. Devanbu. Ownership, experience and defects: A fine-grained study of authorship. In *ICSE*, pages 491–500. ACM, 2011.

[33] F. Rahman and P. Devanbu. How, and why, process metrics are better. In *ICSE*, pages 432–441. IEEE, 2013.

[34] F. Rahman, D. Posnett, and P. Devanbu. Recalling the imprecision of cross-project defect prediction. In *FSE*, page 61. ACM, 2012.

[35] F. Rahman, D. Posnett, I. Herraiz, and P. Devanbu. Sample size vs. bias in defect prediction. In *ESEC/FSE*, pages 147–157. ACM, 2013.

[36] X. Robin, N. Turck, A. Hainard, N. Tiberti, F. Lisacek, J.-C. Sanchez, and M. Müller. pROC: an open-source package for R and S+ to analyze and compare ROC curves. *BMC bioinformatics*, page 77, 2011.

[37] P. J. Rousseeuw and C. Croux. Alternatives to the median absolute deviation. *Journal of the American Statistical Association*, 88(424):1273–1283, 1993.

[38] H. Seo, C. Sadowski, S. Elbaum, E. Aftandilian, and R. Bowdidge. Programmers' build errors: a case study (at Google). In *ICSE*, pages 724–734. ACM, 2014.

[39] J. Śliwerski, T. Zimmermann, and A. Zeller. When do changes induce fixes? *ACM Sigsoft Software Engineering Notes*, 30(4):1–5, 2005.

[40] J. Śliwerski, T. Zimmermann, and A. Zeller. When do changes induce fixes? In *MSR*, pages 1–5. ACM, 2005.

[41] C. Spearman. The proof and measurement of association between two things. *The American Journal of Psychology*, 15:72–101, 1904.

[42] S. Stolberg. Enabling agile testing through continuous integration. In *Agile Conference, 2009. AGILE'09.*, pages 369–374. IEEE, 2009.

[43] T. D. V. Swinscow, M. J. Campbell, et al. *Statistics at square one*. Bmj London, 2002.

[44] J. Tsay, L. Dabbish, and J. Herbsleb. Influence of social and technical factors for evaluating contribution in GitHub. In *ICSE*, pages 356–366. ACM, 2014.

[45] B. Vasilescu, S. van Schuylenburg, J. Wulms, A. Serebrenik, and M. G. J. van den Brand. Continuous integration in a social-coding world: Empirical evidence from GitHub. In *ICSME*, pages 401–405. IEEE, 2014.

[46] B. Vasilescu, Y. Yu, H. Wang, P. Devanbu, and V. Filkov. Quality and productivity outcomes relating to continuous integration in GitHub. In *ESEC/FSE*, pages 805–816. IEEE, 2015.

[47] C. Wohlin, P. Runeson, M. Höst, M. C. Ohlsson, B. Regnell, and A. Wesslén. *Experimentation in software engineering*. Springer, 2012.

[48] H. K. Wright, M. Kim, and D. E. Perry. Validity concerns in software engineering research. In *FSE/SDP Workshop on Future of Software Engineering Research*, pages 411–414. ACM, 2010.

[49] Y. Yu, H. Wang, V. Filkov, P. Devanbu, and B. Vasilescu. Wait for it: Determinants of pull request evaluation latency on GitHub. In *MSR*, pages 367–371. IEEE, 2015.

[50] Y. Yu, H. Wang, G. Yin, and T. Wang. Reviewer recommendation for pull-requests in GitHub: What can we learn from code review and bug assignment? *Information and Software Technology*, 2016. to appear.

[51] J. Zhu, M. Zhou, and A. Mockus. Patterns of folder use and project popularity: A case study of GitHub repositories. In *ESEM*, pages 30:1–30:4. ACM, 2014.

[52] T. Zimmermann, R. Premraj, and A. Zeller. Predicting defects for Eclipse. In *PROMISE*, pages 9–9. IEEE, 2007.